\documentclass[%
 prl,
rsi,%
 amsmath,amssymb,
 reprint,%
]{revtex4-1}

\usepackage{graphicx}
\usepackage{dcolumn}
\usepackage{bm}
\usepackage{xcolor}
\usepackage{soul}
\renewcommand{\hl}[1]{#1}

\begin{document}

\preprint{AIP/123-QED}

\title{Molecular mobility in driven monomeric and polymeric glasses}
\author{J\"org Rottler}
\email{jrottler@physics.ubc.ca}
\affiliation{ Department of Physics and Astronomy and Quantum Matter Institute, University of British Columbia, Vancouver BC V6T 1Z1, Canada
}%

\date{\today}

\begin{abstract}
We show that in monomeric supercooled liquids and glasses that are plastically flowing at a constant shear stress $\sigma$ while being deformed with strain rate $\dot{\epsilon}$, the microscopic structural relaxation time $\tau_{\rm str}$ is given by the universal relation $\sigma/G_\infty\dot{\epsilon}$ with $G_\infty$ a modulus. This equality holds for all rheological regimes from temperatures above the glass transition all the way to the athermal limit, and arises from the competing effects of elastic loading and viscous dissipation. In macromolecular (polymeric) glasses, however, the stress decouples from this relaxation time and $\tau_{\rm str}$ is in fact further reduced even though $\sigma$ rises during glassy strain hardening. We develop expressions to capture both effects and thus provide a framework for analyzing mobility measurements in glassy materials.
\end{abstract}

\pacs{Valid PACS appear here}
\maketitle

Glasses form when their structural relaxation time $\tau_{\rm str}$ exceeds the experimentally accessible observation timescales, and the material falls out of equilibrium. The larger $\tau_{\rm str}$, the smaller the molecular mobility. Predicting the dramatic rise of $\tau_{\rm str}$ with decreasing temperature or increasing packing fraction is a central goal for any theory of the glass transition \cite{berthier_theoretical_2011,ediger_perspective:_2012}. In the glassy state, \hl{following a rapid quench in temperature below the glass transition temperature,} the dynamics slows down further due to physical aging, during which the material slowly recovers towards an eventual equilibrium state. However, relaxation speeds up as soon as the glass is deformed, which must be accounted for by any theory of deformation and plastic flow \cite{falk_deformation_2011,chen_theory_2011,fielding_simple_2012,zou_hybrid_2016,roth_polymer_2016,wisitsorasak_dynamical_2017}. On the most basic level, this acceleration can be understood by realizing that the rate of external driving imposes an upper bound on the relaxation timescale, beyond which all aging processes cease. One therefore expects that the relaxation time varies with the inverse of the deformation rate. This picture, however, is incomplete as it does not consider the role of other parameters such as temperature and stress. In this article, we develop a comprehensive description of accelerated dynamics that extends from supercooled fluids to athermal glasses, and includes all relevant deformation variables.

In soft glasses such as colloidal mixtures, gels and soft pastes, the microscropic structural relaxation time can be accessed through dynamic light scattering \cite{van_megen_measurement_1998,cipeletti_universal_2003}. These materials are usually considered {\em athermal}, i.e. rearrangements are purely due to mechanical excitation. 
For instance, experiments on soft pastes have found that $\tau_{\rm str}$ scales with the visocity of the material under shear \cite{cloitre_glassy_2003}, while $\tau_{\rm str}\propto \dot{\epsilon}^{-1}$ was reported for a colloidal glass sheared at fixed strain rate $\dot{\epsilon}$ \cite{di_leonardo_aging_2005}. Mechanical measurements in similar systems provide further evidence of accelerated dynamics during deformation \cite{agarwal_strain-accelerated_2011}.

Polymer glasses form a second important class of glassy materials with countless technological applications \cite{roth_polymer_2016}. In contrast to the soft glasses, the fundamental building blocks in polymer glasses are molecules, and thermal effects are significant. Both NMR measurements \cite{loo_chain_2000} and fluorescence spectroscopy techniques \cite{lee_dye_2008, lee_direct_2009,hebert_effect_2015,hebert_reversing_2017} have been used to characterize accelerated dynamics at the segmental level \cite{loo_chain_2000}. These experiments confirm that the most important deformation variable that controls the relaxation time is the strain rate $\dot{\epsilon}$. Further insight into accelerated dynamics comes from atomistic molecular dynamics simulations, which report close correlations between the macroscopic deformation and the rate of torsional transitions in glassy polyethylene \cite{capaldi_enhanced_2002} or the mobility of chain segments in polystyrene and polycarbonate \cite{lyulin_strain_2005}. Coarse-grained simulations can access larger ranges of strain rates, and also indicate a $\tau_{\rm str}\propto \dot{\epsilon}^{-1}$ proportionality for different measures of the microscopic relaxation time \cite{riggleman_free_2007,riggleman_nonlinear_2008,warren_microscopic_2010}. Here, we build on this work in order to develop a comprehensive picture of molecular mobility in flowing glasses. We also seek to understand any differences that occur in polymer glasses relative to small molecule or soft glasses that lack chain connectivity. 

Our analysis is based on molecular dynamics simulations of well-studied model amorphous solids. We first consider the Wahnstr\"om model \cite{wahnstrom_molecular-dynamics_1991}, a 50/50 mixture of 3D Lennard Jones (LJ) particles, as a representative model for monatomic glasses. We report all results in reduced simulation units. A periodic simulation box is filled with 100,000 atoms \hl{(at number density $\rho=1.296$)}, equilibrated at {\hl a high} temperature $T=1$ \hl{(where the mixture is in the fluid state)} and then rapidly quenched to lower temperatures varying between $T=0.0-0.7$. Since previous work has established a glass transition temperature of $T\approx 0.46$ for this model \cite{malins_identification_2013}, this temperature range includes both supercooled fluids and the athermal limit. After an aging period of $5,000$ LJ time units \hl{at fixed volume}, we impose uniaxial tensile volume-conserving deformation at constant true strain rates, i.e. $\dot{\epsilon}_z=-2 \dot{\epsilon}_y=-2\dot{\epsilon}_x$, in a range of $10^{-6}<\dot{\epsilon}_z<10^{-3}$, and measure the tensile stress response $\sigma_z-\sigma_x$. Throughout the simulation, the temperature is held constant with a Langevin thermostat \hl{(addition of frictional and random forces)}, and we monitor the monomeric structural relaxation time as a measure of (spatially averaged) inverse molecular mobility via the decay of the self-intermediate scattering function (ISF)
\begin{equation}
F_s(q,t,\epsilon_z)=\langle\exp{[i{\bf q}\cdot\Delta{\bf r}_{NA}(t,\epsilon_z)]}\rangle,
\end{equation}
\hl{where $\epsilon_z$ is the macroscopic (box) strain in the tensile direction and $\Delta {\bf r}_{NA}$ denotes nonaffine displacements, where the trivial motion arising from the global deformation has been subtracted.} We obtain $\tau_{\rm str}(\epsilon_z)$ by fitting $F_s(q,t,\epsilon_z)$ to a stretched exponential (KWW) function $\exp[-(t/\tau_{\rm str}(\epsilon_z))^\beta]$.
By setting $|q|=2\pi$, the ISF decays when the larger particles have moved of order their own diameter, and thus $\tau_{\rm str}$ is the time for particles to break local cages. Light scattering experiments measure directly the ISF, and fluorescence spectroscopy experiments report a similar relaxation time that originates from motion at the segmental level.

\begin{figure}[t]
\includegraphics[width=\linewidth]{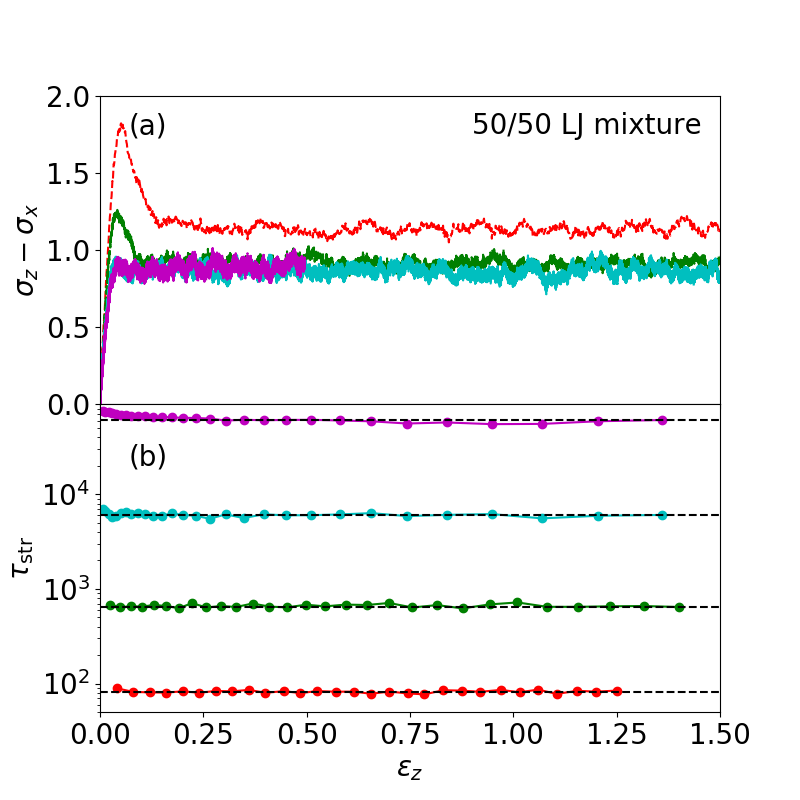}
\caption{(a) True stress vs true strain during deformation at constant true strain rates $10^{-3}$ (red), $10^{-4}$ (green), $10^{-5}$ (cyan), $10^{-6}$ (magenta) and temperature $T=0.3<T_g$. (b) Structural relaxation time $\tau_{\rm str}$ vs.~true strain.}
\label{fig1}
\end{figure}

Figure \ref{fig1}(a) shows representative normal stress differences vs. strain from such simulations, and reveals the typical mechanical response of glassy solids: the stress first rises linearly and then peaks at strains of order $0.1$ before falling back onto a steady state plastic flow plateau. The height of the peak stress and hence the degree of softening depend on the preparation and aging history. We are primarily interested in the post-yield, plastic flow regime, where the flow stress is constant during deformation. Its variation with strain rate and temperature has already been rationalized in previous work \cite{rottler_shear_2003,chattoraj_universal_2010}. The quantity of interest here is the relaxation time $\tau_{\rm str}$ during deformation, displayed in panel (b) of Fig.~\ref{fig1}. In the post-yield plastic flow regime, \hl{$\tau_{\rm pl}\equiv \tau_{str}$} is constant and decreases with increasing strain rate. 

\begin{figure}[t]
\includegraphics[width=\linewidth]{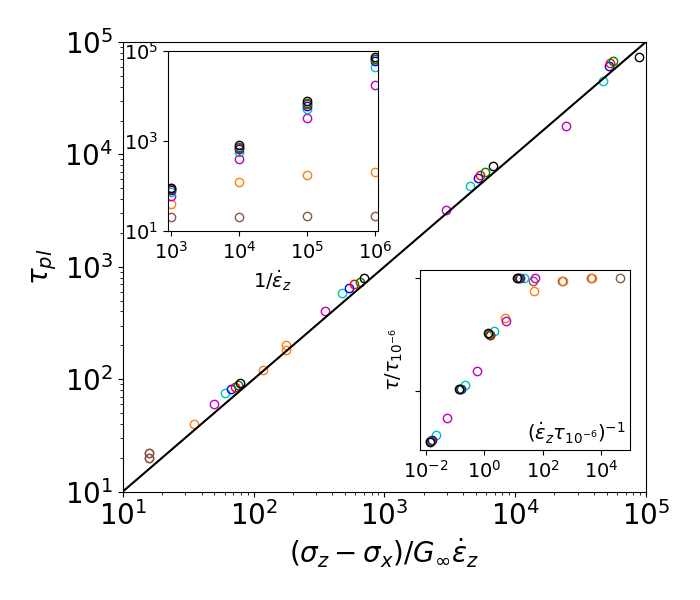}
\caption{Main panel: post-yield relaxation times $\tau_{pl}$ vs. $(\sigma_z-\sigma_x)/G_\infty\dot{\epsilon}_z$, where $G_\infty=17$ was measured in the fluid state. Data for 8 temperatures $T=0.0$ (black), $T=0.1$ (green), $T=0.2$ (red), $T=0.3$ (blue), $T=0.4$ (cyan), $T=0.5$ (magenta), $T=0.6$ (orange), and $T=0.7$ (brown) are shown. Upper inset: same data but plotted against $1/\dot{\epsilon}_z$ only. Lower inset: data collapse when time is measured in units of the longest relaxation time. }
\label{fig2}
\end{figure}

We now proceed with a quantitative analysis of the relaxation time in temperature - strain rate parameter space. The upper inset of Fig.~\ref{fig2} shows $\tau_{pl}$ for 8 different temperatures each as a function of inverse strain rate. We see that the data points from $T<T_g$ are indeed proportional to $\dot{\epsilon}_z^{-1}$, but the prefactor varies with temperature. For $T>T_g$, however, the behavior changes dramatically and $\tau_{\rm pl} \approx {\rm const.}$ at $T=0.7$. We thus observe that the system is crossing over from yield stress fluid (low $T$) via shear thinning to Newtonian (high $T$) behavior. The lower inset shows that the relaxation times can be collapsed onto a common curve if all times are rescaled in terms of the relaxation time measured at the slowest available strain rate. 

Can these trends be captured in a common framework? A starting point is provided by the high temperature fluid, whose viscoelastic behavior is usually well described by phenomenological models introduced by Maxwell 150 years ago: the fluid flows with viscosity $\eta=G_\infty\tau_{M}$, where $G_\infty$ is the elastic modulus in rapid deformation and $\tau_{M}$ is a (fixed) Maxwell relaxation time.  For a Newtonian fluid, both $G_\infty$ and $\tau_{M}$ are material constants and hence $\eta={\rm const.}$ as well. The strain between rearrangements $\dot{\gamma}\tau_M$ thus varies proportional to the shear rate. In the glassy regime, however, the (effective) viscosity \hl{$\eta=(\sigma_z-\sigma_x)/\dot{\epsilon}_z$} depends on shear rate \cite{bonn_laponite:_2002}, and shear thinning with eventual yield stress fluid behavior is found. In contrast to the fluid regime, rearrangements are now driven by the deformation field, and hence the strain between rearrangements is constant. Despite these different mechanisms, we propose that the steady-state rheology at all temperatures is universally governed by the competition between the rates of elastic loading $(G_\infty \dot{\epsilon}_z)$ and viscous dissipation $((\sigma_z-\sigma_x)/\tau)$. Equality of the two terms defines a timescale $\tau$, and we thus suggest

\begin{equation}
\label{mon-eq}
\tau=\tau_{pl}=c(\sigma_z-\sigma_x)/G_\infty\dot{\epsilon}_z.
\end{equation}

The main panel of Fig.~\ref{fig2} tests this prediction and shows that this formula collapses the data for all temperatures onto a single line with slope unity (in general, a unitless constant prefactor $c$ may be needed, but here $c=1$). Remarkably, all temperature effects are now captured by the temperature dependence of the shear stress $\sigma_z-\sigma_x$.  Eq.~\eqref{mon-eq} was found in ref.~\cite{cloitre_glassy_2003} to describe the relaxation times of sheared (athermal) pastes. Our results show that this relationship is far more general, and captures the molecular mobility at finite temperatures all the way to the supercooled liquid regime.  Eq.~\eqref{mon-eq} and its implied data collapse constitutes our first major result. 

In a monomeric or small-molecule glass, the flow stress in steady state is constant. Under these conditions, Fig.~\ref{fig1} evidences that the relaxation time is also constant. Is this still the case when the flow stress depends on strain? This situation occurs in fact in polymer glasses. While their pre-yield behavior is generally accepted to be qualitatively similar to monomeric glasses, polymers exhibit the phenomenon of strain hardening by virtue of their macromolecular character. We now consider a well-known linear bead-spring model \cite{kremer_dynamics_1990}, where monomers interacting with a LJ potential are coupled together with stiff springs to form  chains of $N=500$ beads. These coarse-grained chains have been used before in many simulations of accelerated dynamics \cite{riggleman_free_2007,riggleman_nonlinear_2008,warren_microscopic_2010} and strain hardening \cite{hoy_strain_2007} in glassy polymers .

\begin{figure}[t]
\includegraphics[width=\linewidth]{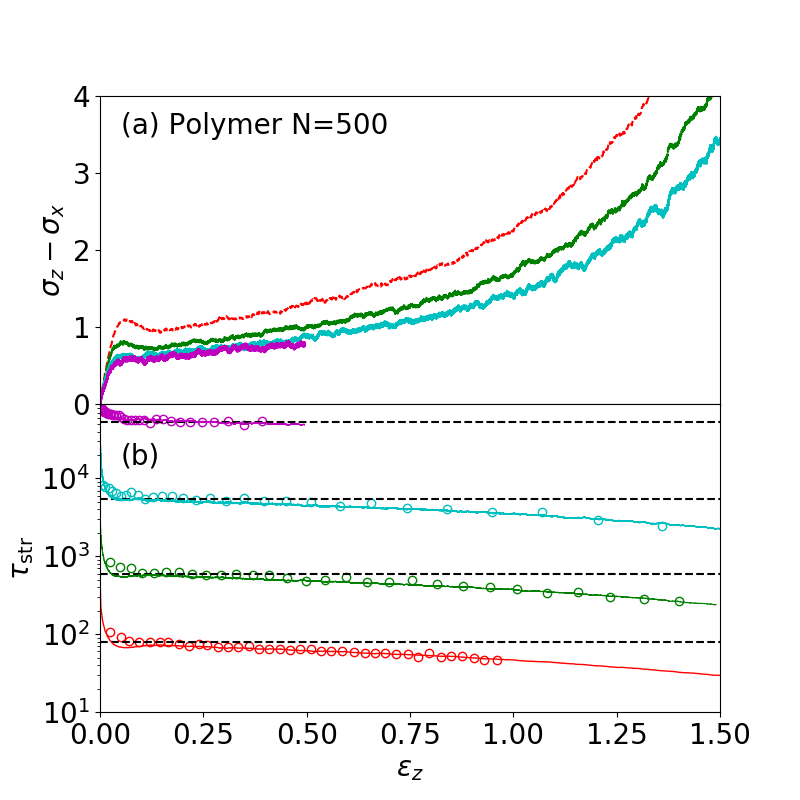}
\caption{(a) True stress vs true strain during deformation of polymer glasses at constant true strain rates  $10^{-3}$ (red), $10^{-4}$ (green), $10^{-5}$ (cyan), $10^{-6}$ (magenta) and temperature $T=0.2$. (b) Structural relaxation time $\tau_{\rm str}$ vs true strain. Solid lines show the functions $\tau_{\rm str}=\tau_0(\sigma_0/(\sigma_z-\sigma_x))^n$ with $n=0.5$.}
\label{fig3}
\end{figure}

As before, we prepare glasses from a rapid quench from the melt and impose deformation after a brief aging period. Since the chains are long enough to be entangled, we use the bond-swap method to equilibrate the chain conformations in the melt \cite{auhl_equilibration_2003}. The stress-strain curves obtained from the same uniaxial deformation at constant true strain rate in Fig.~\ref{fig3}(a) now exhibit substantial hardening. The relaxation times measured again from the decay of the ISF at various deformation strains are shown in Fig.~\ref{fig3}(b). Remarkably, the relaxation times are no longer constant, but decrease monotonically with increasing strain.

The further acceleration of the monomer mobility is unexpected, and runs counter to some theories that predict instead a further slowing down of activated dynamics during hardening \cite{chen_suppressed_2009}. In fact, it also cannot be explained by eq.~\eqref{mon-eq}, from which one would expect an increase of $\tau_{pl}$ with increasing stress. In order to understand the trend observed here, it is useful to recall the molecular mechanisms of strain hardening that were identified by Hoy and Robbins with simulations of the same polymer model \cite{hoy_strain_2007}. They showed that the stress does not arise from loss of conformational entropy of the chains as in rubber elasticity, but reflects instead irreversible work dissipated as heat. Since the chains deform globally affine, the chain connectivity enforces ever increasing plastic activity at the monomer level. Indeed, the rate of plastic rearrangements closely tracks the dissipative component of the stress \cite{hoy_strain_2007}.   

Our finding of accelerating dynamics during hardening is consistent with this picture if we recall that the nonaffine particle displacement is just another characterization of irreversible particle rearrangements. Indeed, Vorselaars et al.~\cite{vorselaars_microscopic_2009} reported that hardening in glassy polymers is closely coupled with an increase in the rate of nonaffine displacements. Noting that  
$\langle \exp[{i{\bf q}\Delta{\bf r}(t,\epsilon_z)}]\rangle\approx \exp[{-{\bf q^2}\langle\Delta_{NA} {\bf r}^2(t,\epsilon_z)\rangle/6}]$ ${\rm if\,\,} {\bf q^2}\langle\Delta_{NA} {\bf r}^2(t,\epsilon_z)\rangle/6\ll 1,$ we see directly that a faster relaxation implies a faster rate of increase of the mean squared nonaffine monomer displacement. This is exactly what happens as the chains become more and more stretched in the glassy matrix.  Since the rate of plastic rearrangements is strongly correlated with the mobility, we expect the hardening process to decrease the relaxation time (inverse mobility) with the inverse of the hardening stress. In Fig.~\ref{fig3}(b) we find the measured times to follow closely the functional form $\tau_{\rm str}=\tau_0(\sigma_0/(\sigma_z-\sigma_x))^n$ with an exponent $n=0.5$ and $\tau_0$ a rate-dependent prefactor. Here $\sigma_0$ denotes the flow stress, which we take as the normal stress difference near $\epsilon_z=0.12$, where the stress differences $\sigma_z-\sigma_x$ exhibit a minimum. 
 
\begin{figure}[t]
\includegraphics[width=\linewidth]{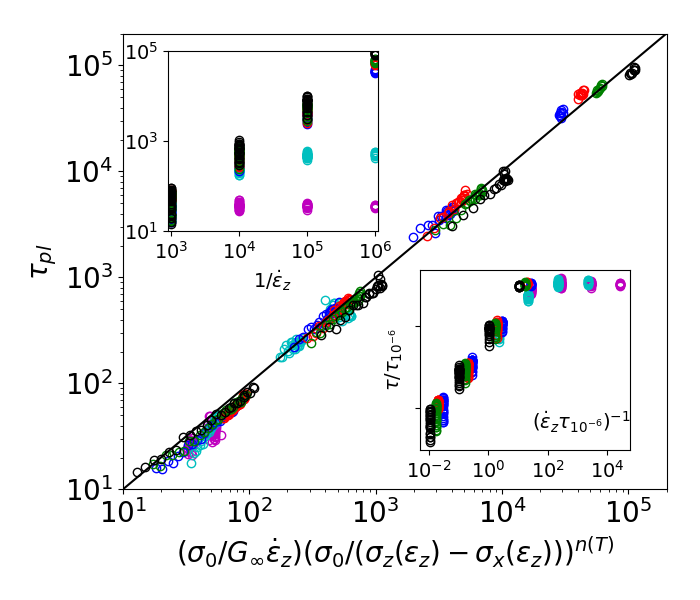}
\caption{Main panel: post-yield relaxation times in polymer glasses, where $G_\infty=11$ and the values of $n= 0.8,0.6,0.5,0.5,0.2,0.05$ at temperatures $T=0, 0.1,0.2,0.3,0.4,0.5$, resp. Colors as in Fig.~\ref{fig2}. Upper inset: same data but plotted against $1/\dot{\epsilon}_z$ only. Lower inset: data collapse when time is measured in units of the longest relaxation time.}
\label{fig4}
\end{figure}

We investigate the relaxation time in the strain hardening regime for multiple temperatures ranging from $T=0.0-0.5$, which exceed the glass transition temperature $T_g\approx 0.35$ for this model \cite{rottler_shear_2003}. In the inset of Fig.~\ref{fig4}, we first observe that for temperatures of $T=0.3$ or lower, the relaxation times overall scale with inverse strain rate. In contrast to the LJ mixture, the times further decrease by several times for a given strain rate, hence the data points no longer overlap. As the temperature increases beyond the glass transition, the relaxation times become independent of rate and also no longer increase in the hardening regime. This further supports the notion that accelerated dynamics is an intrinsically glassy strain hardening effect and absent in entropic hardening that dominates in elastomers and rubbers.

\begin{figure}[t]
\includegraphics[width=\linewidth]{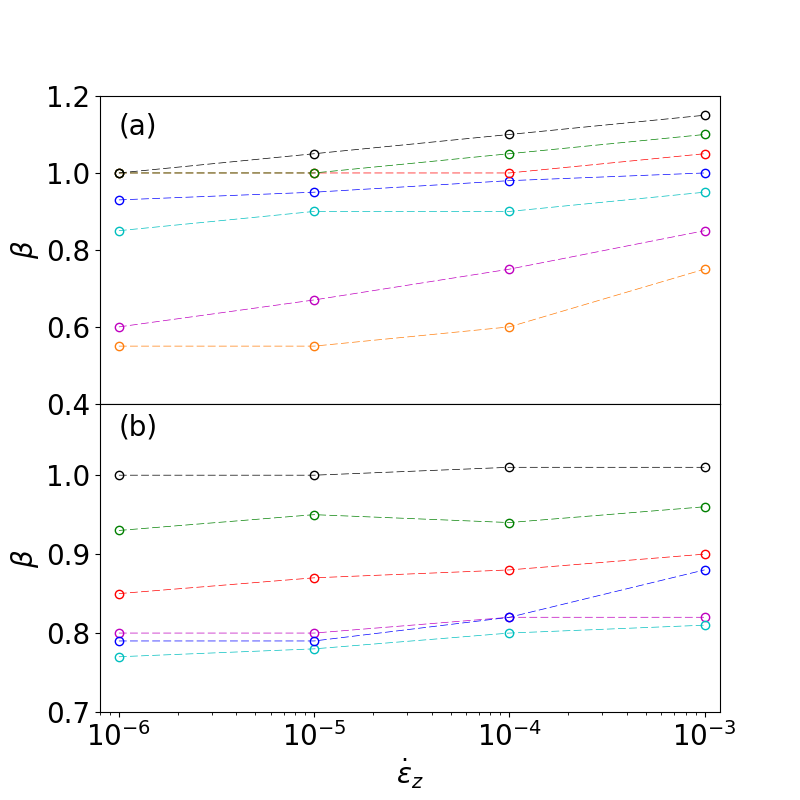}
\caption{KWW-exponent $\beta$ during steady state flow of (a) the binary LJ-mixture and (b) the polymer glass. Colors reflect temperatures $T=0.0$ (black), 0.1 (green), 0.2 (red), 0.3 (blue), 0.4 (cyan), 0.5 (magenta), 0.6 (orange).}
\label{fig5}
\end{figure}

We now introduce a modification of eq.~\eqref{mon-eq} that accounts for the observed trends in polymers. We propose that the effects of strain hardening decouple from the steady state timescale $\tau$ that arises from the balance of elastic loading and viscous dissipation. The latter only sets the prefactor timescale $\tau_0=\sigma_0/G_\infty\dot{\epsilon}_z$ with the flow stress introduced above replacing the steady state shear stress in eq.~\eqref{mon-eq}.  Any further decrease of the relaxation time due to polymeric effects should then be proportional to the hardening stress relative to the flow stress, as evidenced in Fig.~\ref{fig3}(b). We write 
\begin{equation}
\label{pol-eq}
\tau_{pl}=c(\sigma_0/G_\infty\dot{\epsilon}_z)(\sigma_0/(\sigma_z(\epsilon_z)-\sigma_x(\epsilon_z)))^{n(T)},
\end{equation}
where a temperature dependent exponent $n(T)$ is required to interpolate between glassy $(0.5< n < 0.8)$ and elastomeric $(n=0)$ regimes. In the main panel of Fig.~\ref{fig4}, we test this expression against the simulation data and find that it nicely collapses the data from the inset onto a straight line. To achieve this data collapse, we had to optimize the value of the exponent $n$ for each temperature.  Eq.~\eqref{pol-eq} and the implied accelerated dynamics in strain hardening polymers constitute the second major result of the present work. 

Additional information about the flow of the driven glasses can be obtained from the KWW exponent $\beta$, which is fitted to the ISF simultaneously with the relaxation time. We find in Fig.~\ref{fig5} that $\beta$ settles on a steady-state value in the plastic flow regime that increases with shear rate and decreasing temperature. Larger values of $\beta$ correspond to a more homogeneous distribution of relaxation times. In the shear dominated regime at low temperatures, $\beta$ is close to one as the imposed shear rate is the only dynamical timescale in the system. At higher temperatures, thermally activated relaxation competes with deformation-induced plastic events, and the relaxation time distribution broadens. The dynamics in the binary mixture is more heterogeneous than the polymer.

We have developed predictions of the particle scale mobility in the plastic flow regime of supercooled liquids and glasses. The starting point is that the steady state is characterized by a balance of elastic loading and viscous stress relaxation. For a Herschel Bulkley material with the flow stress given by $\sigma_{\rm pl}=\sigma_0+a\sqrt{\dot{\epsilon}}$, eq.~\eqref{mon-eq} predicts $\tau_{\rm pl}=\sigma_0/\dot{\epsilon}+a/\sqrt{\dot{\epsilon}}$, while $\tau_{\rm pl}={\rm const.}$ for a Newtonian fluid.  It is of course expected that the macroscopic viscosity is related to microscopic structural relaxation, but the fact that eq.~\eqref{mon-eq} interpolates universally across all rheological regimes commonly observed in glassy materials has so far not been demonstrated. In an energy landscape viewpoint, the deformation rate sets the global time scale for cage deformation, while the stress further lowers the barriers for cage breaking. In strain hardening polymers, by contrast, accelerated dynamics arises from increased dissipation due to {\em irreversible} chain deformation in the glassy matrix that is no longer balanced by an elastic stress. As a result, the polymeric contribution to the structural relaxation time decouples from the balance between elastic and viscous stresses, and is instead governed by the increased amount of local nonaffine segmental motion as the chains are forced to stretch globally.

Our work has established a robust link between microscopic relaxation time and the macroscopic deformation variables stress and strain rate. The predictions of eqs.~\eqref{mon-eq} and \eqref{pol-eq} are readily testable in soft glasses with light scattering \cite{cloitre_glassy_2003} and in polymer glasses with fluorescence spectroscopy techniques \cite{lee_direct_2009}. Our results also provide an important new benchmark for the formulation of theories of the deformation of glassy solids \cite{wisitsorasak_dynamical_2017} and polymers \cite{chen_theory_2011,fielding_simple_2012,zou_hybrid_2016,roth_polymer_2016}, which frequently invoke a Maxwell material as constitutive law and the evolution of the mobility as an internal state variable.

\begin{acknowledgments}
We thank A.~Nicolas and J.-L.~Barrat for helpful comments on the manuscript and the Natural Sciences and Engineering Research Council of Canada for financial support.
\end{acknowledgments}

%

\end{document}